\documentclass[epj]{svjour}
\usepackage{graphicx}
\usepackage{amsmath}
\usepackage{amssymb}
\usepackage{dcolumn}
\usepackage{bm}
\usepackage[mathlines]{lineno}

\usepackage[utf8]{inputenc}
\usepackage[T1]{fontenc}
\usepackage{mathptmx}

\begin{document}
\title{Sub-bosonic (deformed) ladder operators}
\author{J. Damastor Serafim\inst{1} \and Ricardo Ximenes\inst{2} \and Fernando Parisio\inst{1}
}                     
%
%
\institute{Departamento de
F\'{\i}sica, Universidade Federal de Pernambuco, Recife, Pernambuco
50670-901, Brazil \and Department of Physics, University of Wisconsin-Madison, Madison, WI 53706, USA.}
\date{Received: date / Revised version: date}
%
\abstract{
The canonical operator $\hat{a}^{\dagger}$  ($\hat{a}$) represents the ideal process of adding (subtracting) an {\it exact} amount 
of energy $E$ to (from) a physical system in both elementary quantum mechanics and quantum field theory. 
This is a ``sharp'' notion in the sense that no variability around $E$ is possible at the 
operator level. In this work, we present a class of deformed creation and annihilation operators that originates from a rigorous notion of fuzziness.
This leads to deformed, sub-bosonic commutation relations inducing a simple algebraic structure with modified eigenenergies and Fock states.
In addition, we investigate possible consequences of the introduced formalism in quantum field theories, as for instance, deviations from linearity in the
dispersion relation for free quasibosons. 
\PACS{
      {PACS-key}{discribing text of that key}   \and
      {PACS-key}{discribing text of that key}
     } 
} 
\maketitle
\section{Introduction}
\label{intro}
While quantum field theory (QFT) is at the heart of our understanding of a large variety of phenomena, 
there remains some discrepancies between theoretically predicted and observed quantities.  
Perhaps, the most known of these problems is related to the zero-point energy that appears as a result of field quantization \cite{weinberg,peskin}. 
On the one hand, it provides the conceptual framework to explain the finite lifetimes of excited atoms and the Casimir effect, 
for instance, while, on the other hand, it leads to an abyss between observed and predicted values of the cosmological 
constant \cite{rugh,padma}. One possible path to deal with these and other issues in QFT is the more or less ad hoc introduction
of deformed field operators. An illustration of the consequences of such a strategy is the natural emergence of
a minimal spatial scale in prospective quantum theories of gravity employing slightly deformed position and
momentum operators \cite{kempf,kempf2,kempf3}.

In a different context, deformed operators in quantum mechanics have been addressed by Green, back in 1953 \cite{green},
as a way to investigate hypothetical particles whose collective behavior would be somewhere between the Fermi-Dirac and 
Bose-Einstein statistics (see also \cite{hawton}). Since then, the study of deformed-operator algebras is motivated by a variety of reasons
ranging from more formal/aesthetic ones \cite{macfarlane,biedenharn,fu} to attempts to cope with foundational 
physical problems not only in QFT but also in elementary quantum mechanics.  An example of the latter is the translation operator for systems 
with a position-dependent effective mass (as it may be the case of electrons in semi-conductors) leading to a deformed position-momentum 
commutation relation \cite{raimundo}, which, in turn, provides a surprising foundation for the, otherwise empirical Morse potential \cite{raimundo2}.

In this work, we adopt a distinct route which also results in a class of deformed commutation relations for modified creation and annihilation operators.
Ladder operators can be seen as representing the ideal processes of exchanging a precise amount of energy. We introduce
some degree of variability in these processes, at the operator level, by providing a rigorous definition of sub-bosonic ladder operators which embody
an average over frequencies, so that, the originally sharp operations become ``blurred''. After setting the general formalism, we investigate sub-bosonic harmonic 
oscillators and the corresponding deformed eigenfunctions, symmetry operations, coherent states, as well as, some potential consequences in QFT, 
such as nonlinear dispersion relations for free excitations. The terminology ``sub-bosonic'' will be justified shortly.

\section {Preliminaries: Ladder operators}
\label{pre}
We start by addressing the elementary situation in which ${\cal H}$ is the Hilbert space associated with a particle of mass $m$ in one spatial dimension. 
Consider two operators $\hat{A}=\alpha \hat{q}+i\beta \hat{p}$ and $\hat{A}'=\alpha'\hat{q}+i\beta'\hat{p}$, where $\hat{q}$ and $\hat{p}$ 
are the canonically conjugated position and momentum operators acting on ${\cal H}$. It is then immediate that 
\begin{equation}
\label{commut_gen}
[\hat{A},\hat{A}']=\hbar(\beta \alpha'-\alpha\beta').
\end{equation}
If we set 
\begin{equation}
\nonumber
\alpha=\frac{1}{\sqrt{2}b}\;\;\; \mbox{and}\;\;\; \beta=\frac{b}{\sqrt{2}\hbar}, \;\;\; \mbox{with} \;\;\;b=\sqrt{\frac{\hbar}{m\omega}},
\end{equation}
$1/\omega$ being a constant with dimension of time, $\hat{A}$ formally coincides with the annihilation operator,
\begin{equation}
\label{a}
\hat{A}\equiv \hat{a}_{\omega}=\frac{1}{\sqrt{2}}\left( \frac{\hat{q}}{b}+i\frac{b\hat{p}}{\hbar}\right)=\sqrt{\frac{m\omega}{2\hbar}}\hat{q}+\frac{i}{\sqrt{2\hbar m \omega}}\hat{p},
\end{equation}
related to a harmonic potential of angular frequency $\omega$. 
If we make analogous assignments for $\alpha'$, $\beta'$ and $b'$, related to $\omega'$, then we get the annihilation operator $\hat{a}_{\omega'}$ associated 
with a harmonic potential of angular frequency $\omega'$. 

It is important to note that we can make these definitions whether or not the system is, in fact, subjected to a harmonic potential. We can simply see $\hat{a}_{\omega}$ and $\hat{a}_{\omega'}$ as different linear combinations of $\hat{q}$ and $\hat{p}$. Importantly, in this context, $\hat{a}_{\omega}$ and $\hat{a}_{\omega'}$ act on the {\it same} Hilbert space. It is then a trivial exercise to show that (\ref{commut_gen}) assumes the curious form
\begin{equation}
\label{commut_1}
[\hat{a}_{\omega},\hat{a}_{\omega'}]=\frac{1}{2}\left(  {\frac{\sqrt{\omega'}}{\sqrt\omega}}-{\frac{\sqrt\omega}{\sqrt{\omega'}}}\right),
\end{equation}
which is non-vanishing in general (except when $\omega'=\omega$) and depends only on the ratio of the angular frequencies.
This should be contrasted with the relation $[\hat{a}_{\omega},\hat{a}_{\omega'}]=0$, when the operators refer to quantized bosonic fields, for which $\hat{a}_{\omega}$ and $\hat{a}_{\omega'}$ concern distinct modes, thus, acting on {\it different} Hilbert spaces. 

Back to elementary quantum mechanics, by eliminating $\hat{q}$ and $\hat{p}$ in the expressions for $\hat{a}_{\omega}$ and $\hat{a}_{\omega'}$ one obtains
\begin{equation}
\label{a2}
\hat{a}_{\omega+\Delta \omega}=\frac{1}{\sqrt{1+\Delta \omega/\omega}}\left[\hat{a}_{\omega}+\frac{\Delta \omega}{2\omega}(\hat{a}_{\omega}^{\dagger}+\hat{a}_{\omega})\right],
\end{equation}
where, for convenience, we set $\Delta \omega\equiv\omega'-\omega$ and an analogous expression holds for $\hat{a}_{\omega+\Delta \omega}^{\dagger}$. Therefore, one can express the annihilation associated with an arbitrary frequency as a linear combination of annihilation and creation processes of any fixed, distinct frequency. Of course, for $\Delta \omega=0$ we simply get $\hat{a}_{\omega}$.The formal dependence of ladder operators on the frequency, crucial to the remainder of this work, has not been much investigated in the literature. One exception is its use to define phase-space path integrals with coherent states related to variable angular frequencies \cite{parisio}

\section {Sub-bosonic operators}
\label{fuzzy}
Usually, we would restrict the validity of the previous relations (\ref{a}), (\ref{commut_1}), and (\ref{a2}) to $\omega, \omega' >0$  ($\Delta \omega\ge -\omega$). However, we note that the formal replacement $\omega \rightarrow - \omega$ takes annihilation into creation (and vice-versa):
$$\hat{a}_{\omega} \rightarrow \hat{a}_{-\omega}=i\hat{a}^{\dagger}_{|\omega|}$$ 
in definition (\ref{a}). We must be careful about the branch of $\sqrt{-1}$ to be considered. Throughout this work we will adopt $\sqrt{-1}=i$ (accordingly $1/\sqrt{-1}=-i$), leading to $[\hat{a}_{\omega},\hat{a}^{\dagger}_{|\omega|}]=-i[\hat{a}_{\omega},\hat{a}_{-\omega}]$. But, from (\ref{commut_1}) we get $[\hat{a}_{\omega},\hat{a}_{-\omega}]=(i-1/i)/2=i$, therefore $[\hat{a}_{\omega},\hat{a}^{\dagger}_{|\omega|}]=1$, which reproduces the expected result for $\omega>0$. With these extensions in mind, expression (\ref{a2}) becomes well defined for $\Delta \omega \in (-\infty, +\infty)$, which is a fundamental point for the developments to be presented in the remainder of this work.

The particular kind of deformation we intend to introduce here relates to the fact that creation (annihilation) operators can be seen as the mathematical representations of the ideal process of adding (subtracting) the exact amount of energy $E_{\omega}=\hbar \omega$ to (from) a harmonic system. This is a ``sharp'' definition in the sense that no fluctuations around $E_{\omega}$ are possible at the operator level. Consider that some, unspecified factors make the physical processes inducing creation (annihilation) bear some level of variability. This naturally leads to the definition of averaged operators
\begin{equation}
\label{fuzzy}
\bar{a}_{\omega}=\int_{-\infty}^{+\infty}f(\Delta \omega) \hat{a}_{\omega+\Delta \omega} \,d\Delta\omega,
\end{equation}
with an analogous definition being valid for $\bar{a}_{\omega}^{\dagger}$. The function $f(\Delta \omega)$ is a normalized, real-valued distribution peaked at $\Delta \omega=0$ with dimension of
time (inverse frequency) and $ \hat{a}_{\omega+\Delta \omega}$ is given by (\ref{a2}), which makes the above ``operator integral'' be mapped into a pair of complex valued, standard integrals. In the case of a Dirac-delta distribution, $f(\Delta \omega)=\delta(\Delta \omega)$, we get the usual creation and annihilation operators. 
Explicitly, we get
\begin{equation}
\label{fuzzya}
\bar{a}_{\omega}=(I_0+I_1/2)\hat{a}_{\omega}+(I_1/2)\hat{a}_{\omega}^{\dagger},
\end{equation}
and
\begin{equation}
\label{fuzzyad}
\bar{a}^{\dagger}_{\omega}=(I^{*}_0+I^{*}_1/2)\hat{a}_{\omega}^{\dagger}+(I^{*}_1/2)\hat{a}_{\omega},
\end{equation}
where
\begin{equation}
\label{integr}
I_k=\int^{\infty}_{-\infty}\frac{x^kf'(x)}{\sqrt{1+x}} dx=\int^{\infty}_{-\infty}\frac{g_k(x)}{\sqrt{x}} dx,
\end{equation}
with $x=\Delta \omega/\omega$, $f'(x)=\omega f(\Delta \omega)$, $g_k(x)= (x-1)^kf(x-1)$, and $k=0,1$. Therefore we see that the deformed creation (annihilation) operators can be represented simply as a linear combination of $\hat{a}_{\omega}$ and $\hat{a}_{\omega}^{\dagger}$. However, it is important to note that this linear combination not necessarily corresponds to a Bogolyubov transformation, since, as we will see next (see section \ref{FF}), $[\bar{a}_{\omega},\bar{a}_{\omega}^{\dagger}]\le1$, in general. 

The more concentrated is $f(x)$ around $x=0\, (\Delta \omega=0)$ the smaller is $I_1/2$ in comparison with $I_0$. In other words, for weak deformation, $\bar{a}_{\omega}$ is a combination of a large proportion of sharp annihilation with a small amount of sharp creation.
We now turn our attention to the non-canonical commutation relations. It is easy to show that
\begin{equation}
\label{comm2}
[\bar{a}_{\omega},\bar{a}_{\omega}^{\dagger}]=|I_0|^2+\Re{(I_0I^{*}_1)}\equiv {\cal C},
\end{equation}
which is always a real number and characterizes the deformed algebra underlying the sub-bosonic operators. We call $ {\cal C}$ the commutation function which, as we will see, may depend on the frequency $\omega$ and on the characteristic width $\Gamma$ of the distribution $f(\Delta \omega)$. The above, deformed commutation relation is arguably the simplest variation of the canonical relation, since it is still proportional to the identity operator. Compare, for instance, with the $q$-deformed bosonic realizations of the quantum groups SU$(n)_q$ \cite{macfarlane,biedenharn,fu}. 
In spite of this simplicity, the sub-bosonic commutation relations may originate non-trivial physical structures, particularly in QFT. We now provide a quite general result which we state as follows.

{\bf Proposition 1}: Let $g_k(x)$ in equation (\ref{integr}) be a function satisfying the two requirements: (i) it is analytic in the upper half of the complex plane including the real line (${\rm H} \cup \mathbb{R}$), except, possibly for a finite number of removable singularities in ${\rm H}$; and (ii) $g_{k}(Re^{i\theta}) \underset{R \to \infty}{\sim} R^\alpha$, $g_{k}(\epsilon e^{i\theta}) \underset{\epsilon \to 0}{\sim} \epsilon^\beta$ with $\alpha<-1/2$ and $\beta>-1/2$. Under these conditions, the explicit expressions for the sub-bosonic annihilation operator is:
\begin{equation}
\label{prop1}
\bar{a}_{\omega}=2 \pi i \left[\sum_{z=a}\frac{\operatorname{Res}\,g_{0}(z)}{\sqrt{z}}\hat{a}_{\omega}\right.+\sum_{z=a}\left.
\frac{\operatorname{Res}\,g_{1}(z)}{2\sqrt{z}}(\hat{a}_{\omega}+\hat{a}_{\omega}^{\dagger}) \right],
\end{equation}
where $a$ represents the set of removable singularities in $H$ and ``Res'' denotes residue of $g_k$. A completely analogous expression holds for $\bar{a}_{\omega}^{\dagger}$. The proof of this proposition is given in the appendix.
\section{Examples and Developments}
\label{ED}
\subsection{Natural-line width distribution}
\label{NL}
Perhaps the most compelling distributions to consider are those related to natural line widths. In other words, we consider a natural width at the ladder-operator level by attributing to $f$ a Lorentzian dependence:
\begin{equation}
\label{lor}
f(\Delta \omega)= \frac{1}{\pi}\frac{\Gamma/2}{\Delta \omega^2+(\Gamma/2)^2}
\end{equation}
where $\Gamma \ge 0$ characterizes the width of the distribution, which may depend on the frequency, $\Gamma=\Gamma(\omega)$. In terms of the dimensionless parameter $\zeta=\Gamma/2\omega$ and variable $x=\Delta \omega/\omega$ we have
\begin{equation}
 f'(x)=\omega f(\Delta \omega)= \frac{1}{\pi} \frac{\zeta}{x^2+\zeta^2}.
\end{equation}
Using this expression in (\ref{fuzzy}) and setting $g_{k}(x)=(x-1)^k f'(x-1)$, we get
\begin{equation}
g_{k}(x)=\frac{1}{\pi} \frac{(x-1)^k\zeta}{(x-1)^2+\zeta^2}=\frac{1}{\pi} \frac{(x-1)^k\zeta}{(x-z_{+})(x-z_{-})},
\end{equation}
with $k=0,1$. Here we have defined $z_{\pm}=1 \pm i\zeta$. We can see that both $g_{0}$ and $g_{1}$ are analytic, except for the simple poles at $z_{\pm}$. Thus, condition (i) is satisfied. Now, let us look at condition (ii). The coefficients $\alpha$ and $\beta$ for $g_{0,1}$ are given by
\begin{eqnarray}
g_{0}(Re^{i\theta}) \underset{R \to \infty}{\to} R^{-2} \Rightarrow \alpha=-2 < -1/2,\\
 g_{1}(\epsilon e^{i\theta}) \underset{\epsilon \to 0}{\to} const.=\epsilon^{0} \Rightarrow \beta=0 > -1/2.
\end{eqnarray}
\begin{eqnarray}
g_{0}(Re^{i\theta}) \underset{R \to \infty}{\to} R^{-1} \Rightarrow \alpha=-1 < -1/2, \\
 g_{1}(\epsilon e^{i\theta}) \underset{\epsilon  \to 0}{\to} const.=\epsilon^{0} \Rightarrow \beta=0 > -1/2.
\end{eqnarray}
Therefore, condition (ii) is also fulfilled. We now write the residue of $g_{0,1}(z)$ at $z_{+} \in H$ as
\begin{equation}
\underset{z=z_{+}}{\operatorname{Res}}~\frac{g_{k}(z)}{\sqrt{z}}= \underset{z=z_{+}}{\lim}(z-z_{+})\frac{g_{k}(z)}{\sqrt{z}}=\frac{1}{2\pi i }\frac{(i\zeta)^k}{\sqrt{1+i\zeta}}.
\end{equation}
Hence, for the Lorentzian distribution 
\begin{equation}
\bar{a}_{\omega}= \left( \frac{1+i\zeta/2}{\sqrt{1+i\zeta}}\right) \hat{a}_{\omega}+\frac{i\zeta/2}{\sqrt{1+i\zeta}}\hat{a}^{\dagger}_{\omega}, 
\end{equation}
and 
\begin{equation}
\bar{a}^{\dagger}_{\omega}= \left( \frac{1-i\zeta/2}{\sqrt{1-i\zeta}}\right) \hat{a}^{\dagger}_{\omega}-\frac{i\zeta/2}{\sqrt{1-i\zeta}}\hat{a}_{\omega}.
\end{equation}
The commutation relation assumes the fairly simple form:
\begin{equation}
[\bar{a}_{\omega},{\bar{a}_{\omega}^{\dagger}}]=\frac{1}{\sqrt{1+\zeta(\omega)^2}},
\end{equation}
where we made the dependence on the angular frequency explicit, $\zeta(\omega)=\Gamma(\omega)/2\omega$. For
any finite $\omega$ and $\Gamma=0$ ($\zeta=0$), we obtain the canonical commutation relation, while, for $\Gamma
\rightarrow \infty$ ($\zeta \rightarrow \infty$), we get $[\bar{a}_{\omega},{\bar{a}_{\omega}^{\dagger}}]\sim \zeta^{-1}\rightarrow 0$.
\subsection{Uniform distribution}
\label{Ud}
Of course, the definition of sub-bosonic operators holds for distributions which are not encompassed by conditions under which Proposition 1 is valid. As an example let us consider the simplest non-trivial realization of Eq. (\ref{fuzzy}),  that is, a uniform distribution over a finite, symmetric interval.
This amounts to $f(\Delta \omega)=1/\Gamma$ for $\Delta \omega \in [-\Gamma/2,\Gamma/2]$, with $f(\Delta \omega)=0$ elsewhere, 
leading to the simple integrals
\begin{equation}
I_k=\omega \int^{+\Gamma/2\omega}_{-\Gamma/2\omega}\frac{x^k\,dx}{\sqrt{1+x}}=\omega \int^{+\zeta}_{-\zeta}\frac{x^k\,dx}{\sqrt{1+x}},
\end{equation}
with $x=\Delta \omega/\omega$, $k=0,1$. These integrals can be readily done but result in quite lengthly expressions for $\bar{a}_{\omega}$ and ${\bar{a}_{\omega}^{\dagger}}$.
We only write below the commutation relation for the sub-bosonic creation and annihilation operators:
\begin{equation}
\label{commU}
   [\bar{a}_{\omega},\bar{a}_{\omega}^{\dagger}]=\frac{2}{3}
  \begin{cases}
   2+\zeta^{-2}\left( \sqrt{1-\zeta^2}-1 \right)  & \text{if $\zeta \le 1$},\\
\zeta^{-1} & \text{if $\zeta > 1$}.
    \end{cases}
\end{equation}
which goes to 1 as $\Gamma \rightarrow 0$, that is $\bar{a}_{\omega} \rightarrow a_{\omega}$, as it should be. Note that, as in the case of the Lorentzian distribution, we obtain
 $[\bar{a}_{\omega},{\bar{a}_{\omega}^{\dagger}}]\sim \zeta^{-1}\rightarrow 0$ as $\zeta \rightarrow \infty$. For both distributions we get $0<[\bar{a}_{\omega},\bar{a}_{\omega}^{\dagger}]\le 1$,
 which represent  specific sub-bosonic commutation relations.
\subsection{Sub-bosonic Harmonic Oscillators and Fock States}
\label{FF}
Consider a simple harmonic oscillator, whose Hamiltonian reads $\hat{H}=\hat{p}^2/2m+m\omega^2\hat{q}^2/2=\hbar \omega(\hat{a}_{\omega}^{\dagger}\hat{a}_{\omega}+\hat{a}_{\omega}\hat{a}_{\omega}^{\dagger})$. This is as far as we can go without using the canonical relation $[\hat{a}_{\omega},\hat{a}^{\dagger}_{\omega}]=1$, that allows us to write $\hat{H}=\hbar \omega\hat{a}_{\omega}^{\dagger}\hat{a}_{\omega}+\hbar \omega/2$. Let us consider, as a working hypothesis, oscillators that have some builtin fuzziness, with $\hat{a}$ ($\hat{a}^{\dagger}$) replaced by $\bar{a}$ ($\bar{a}^{\dagger}$). This hypothesis may be considered an attempt to model the effect of the vacuum of {\it other }fields in the context of QFT. Thus, we write
\begin{equation}
\nonumber
\bar{H}=\hbar \omega(\bar{a}_{\omega}^{\dagger}\bar{a}_{\omega}+\bar{a}_{\omega}\bar{a}_{\omega}^{\dagger}),
\end{equation}
that, due to (\ref{comm2}), becomes the deformed energy observable
\begin{equation}
\nonumber
\bar{H}=\hbar \omega\bar{a}_{\omega}^{\dagger}\bar{a}_{\omega}+\frac{\hbar \omega}{2} \mathcal{C}.
\end{equation}
The commutation relation (\ref{comm2}) induces the simple algebraic structure that we describe in what follows. Let us define the sub-bosonic Fock states $|\bar{n}\rangle$,
such that $\bar{H}|\bar{n}\rangle=\bar{E}_n|\bar{n}\rangle$ .
We demand the operators $\bar{a}_{\omega}$ and $\bar{a}_{\omega}^{\dagger}$ to be genuine annihilation and creation operators. This can be achieved only if the sub-bosonic ``number''  operator is defined by
\begin{equation}
\label{number}
\bar{N}_{\omega}\equiv \mathcal{C}^{-1}\,\bar{a}_{\omega}^{\dagger}\bar{a}_{\omega}.
\end{equation}
With this, we indeed obtain the usual commutators $[\bar{a}_{\omega},\bar{N}_{\omega}]=\bar{a}_{\omega}$ and  $[\bar{a}_{\omega}^{\dagger},\bar{N}_{\omega}]=-\bar{a}_{\omega}^{\dagger}$ and, thus,
\begin{equation}
\nonumber
\bar{a}_{\omega}|\bar{n}\rangle=\sqrt{n}\,\left| \right.\overline{n-1}\left. \right\rangle, \, \bar{a}_{\omega}^{\dagger}|\bar{n}\rangle=\sqrt{n+1}\,\left| \right.\overline{n+1}\left. \right\rangle,
\end{equation}
as intended. Therefore, in terms of $\bar{N}_{\omega}$ we get 
$$\bar{H}=\hbar\omega\mathcal{C}(\bar{N}_{\omega}+1/2).$$ 
It is instructive to determine the deformed fundamental state in terms of the usual Fock states, $|\overline{0}\rangle=\sum_j \alpha_j|j\rangle$. The coefficients $\alpha_j$ can be determined through relation $\bar{a}_{\omega}|\overline{0}\rangle=0$ together with (\ref{fuzzya}), leading to
\begin{equation}
\nonumber
(2I_0+I_1)\sum_{j=1}^{\infty}\alpha_j\sqrt{j}|j-1\rangle+I_1\sum_{j=0}^{\infty}\alpha_j\sqrt{j+1}|j+1\rangle=0,
\end{equation}
which can be rewritten as
\begin{equation}
\sum_{n=1}^{\infty}\left[(2I_0+I_1)\alpha_{n+1}\sqrt{n+1}+I_1\alpha_{n-1}\sqrt{n}\right]|n\rangle=0
\end{equation}
{\it and} 
\begin{equation}
\alpha_1=0.
\end{equation}
This corresponds to a recursion relation whose solution is 
\begin{eqnarray}
\nonumber
\alpha_{2k}&=&(-1)^k\left( \frac{I_1}{2I_0+I_1}\right)^k\sqrt{\prod_{\ell=1}^{k}\left( \frac{2\ell-1}{2\ell}\right)}\,\alpha_0\\
&=&\frac{(-1)^k}{ \pi^{1/4} } \left( \frac{I_1}{2I_0+I_1}\right)^k\sqrt{\frac{(k-1/2)!}{k!} }\,\alpha_0,
\label{alpha}
\end{eqnarray}
with $k=0,1,2,\dots$ and $\zeta=\Gamma/2\omega$, and $\alpha_{n}=0$ for $n$ odd. With this, the original, well-defined parity of the eigenstates remains unchanged. 
Finally, the constant factor is determined, up to a global phase factor, by $\langle \overline{0} |\overline{0}\rangle=1$.

Let us consider the specific example of the Lorentzian distribution. In this case we have
\begin{equation}
\alpha_{2k}=\frac{1}{ \pi^{1/4} } \left( \frac{\zeta}{2i-\zeta}\right)^k\sqrt{\frac{(k-1/2)!}{k!} }\,\alpha_0,
\end{equation}
which together with the normalization condition leads to
\begin{equation}
\label{vac}
|\overline{0}\rangle=\left( \frac{4}{\zeta^2+4} \right)^{1/4}\sum_{k=0}^{\infty} \left( \frac{\zeta}{2i-\zeta}\right)^k\sqrt{\frac{(k-1/2)!}{k!} }\,|2k\rangle.
\end{equation}
So, 
\begin{equation}
\langle 0|\overline{0}\rangle=\alpha_0=\left( \frac{4}{\zeta^2+4} \right)^{1/4},\; |\langle 0|\overline{0}\rangle|^2=\frac{2}{\sqrt{\zeta^2+4}},
\end{equation}
\begin{figure*}
\centering
\resizebox{0.45\textwidth}{!}{
 \includegraphics{{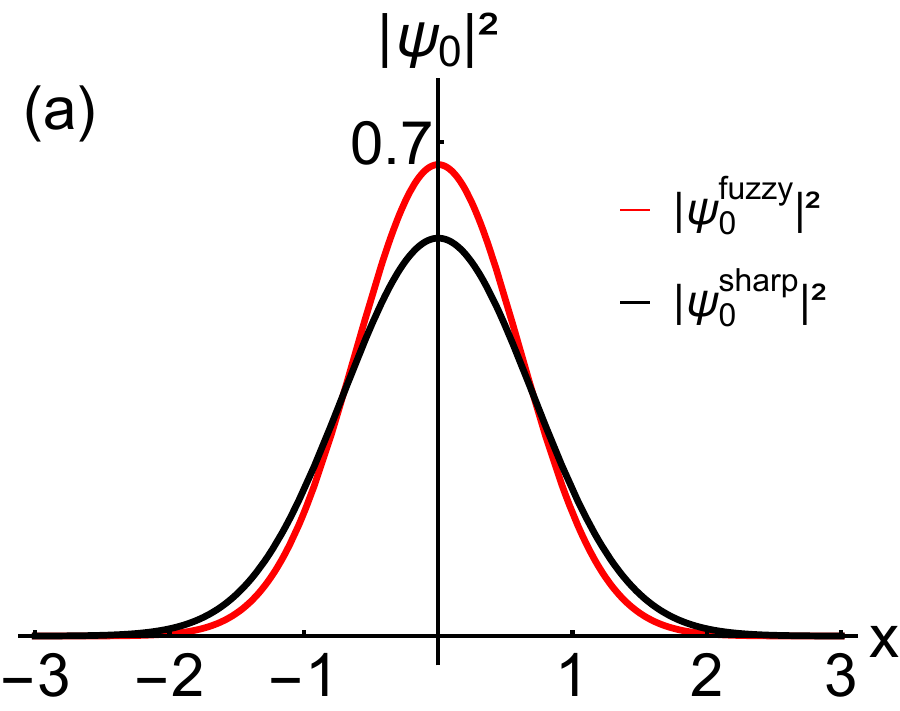}}
}
\resizebox{0.45\textwidth}{!}{
 \includegraphics{{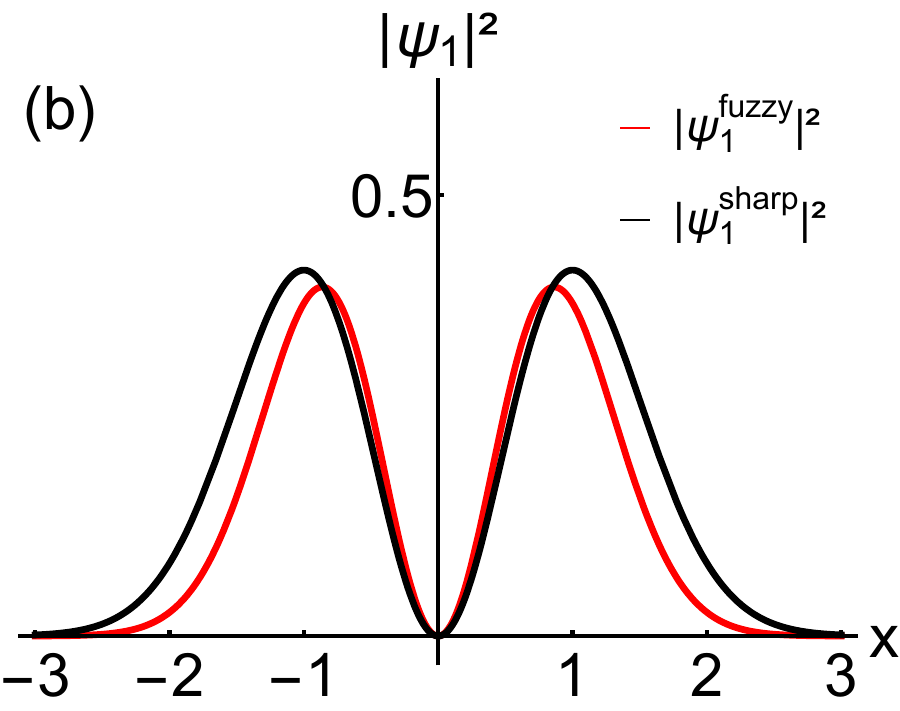}}
}
\caption{(a) Standard (black) and sub-bosonic or fuzzy (red) probability densities for the ground state wave function. The sub-bosonic $\psi$ concerns the Lorentzian distribution
with $\zeta=0.3$. (b) Same as in (a), but for the first excited states. The parity of the states remains unchanged.}
\label{f1}
\end{figure*}
which shows that part of the sub-bosonic lower state does not correspond to the usual ground state, since $\alpha_0 \le 1$. In this exploratory scenario part of the energy that we would attribute to the ``vacuum'' would be distributed in the sharp-field excitations. Any excitation of the sub-bosonic field can be obtained in terms of standard Fock states with the help of equation (\ref{vac}),
see Fig. 1 for a comparison between the spatial probability densities $|\langle x|\overline{n}\rangle|^2\equiv |\Psi^{\rm fuzzy}_n(x)|^2$ and $|\langle x|{n}\rangle|^2\equiv |\Psi^{\rm sharp}_n(x)|^2$ for $n=0,1$. As we will see next, the fact that the sharp and sub-bosonic eigenstates have the same, well-defined parity is a particular instance of a more general result.
To finish this subsection we note that the probability associated with vacuum fluctuations is given by $|\langle \overline{0}| [\bar{a}_{\omega},\bar{a}_{\omega}^{\dagger}]|\overline{0}\rangle|^2={\cal C}^2$. So, given this probabilistic interpretation, we must have $0\le {\cal C}^2 \le 1$, which justifies the terminology ``sub-bosonic''.
\subsection{Symmetries}
According to Wigner's theorem \cite{wigner}, a symmetry transformation acting on the Hilbert space can either be linear and unitary (LU), or antilinear and antiunitary (AA). Now, suppose we know how the sharp operators behave under the symmetry transformation $\mathcal{O}$, which can be of the type LU or AA. We want to investigate how the sub-bosonic operators behave under such transformations. 
From Eq. (\ref{fuzzya}) and for a LU $\mathcal{O}$, we have
\begin{equation}
\mathcal{O}\bar{a}_{\omega}\mathcal{O}^{-1}=\left(I_{0}+\frac{I_{1}}{2}\right)\mathcal{O}\hat{a}_{\omega}\mathcal{O}^{-1}+\frac{I_{1}}{2}\mathcal{O}\hat{a}^{\dagger}_{\omega}\mathcal{O}^{-1}
\end{equation}
Supposing that the sharp operators behave as $\mathcal{O}\hat{a}_{\omega}\mathcal{O}^{-1}=\alpha\hat{a}_{\omega}$ ($\alpha \in \mathbb{R}$) under the transformation, we can see that
\begin{equation}
\mathcal{O}\hat{a}_{\omega}\mathcal{O}^{-1}=\alpha\hat{a}_{\omega} \Rightarrow \mathcal{O}\bar{a}_{\omega}\mathcal{O}^{-1}=\alpha\bar{a}_{\omega}.
\end{equation}
In particular, if $\alpha=1$, we have that 
\begin{equation}
[\mathcal{O},\hat{a}_{\omega}]=0 \Rightarrow [\mathcal{O},\bar{a}_{\omega}]=0.
\end{equation}
This situation is particularly interesting for, suppose we have a system whose Hamiltonian $H$ is a function of the sharp operators only, and that $\mathcal{O}\hat{a}_{\omega}\mathcal{O}^{-1}=\alpha\hat{a}_{\omega}$, $\alpha \in \mathbb{R}$. Then, it is easy to see that if $H$ is invariant under $\mathcal{O}$, the Hamiltonian obtained by substituting the sharp operators by sub-bosonic operators, denoted by $\bar{H}$, will also be invariant under $\mathcal{O}$. As an example, suppose we consider a sub-bosonic Harmonic oscillator. The action of the parity operator in the sharp operator reads $\Pi\hat{a}_{\omega}\Pi^{-1}=-\hat{a}_{\omega}$, and, as we know, the Hamiltonian is parity invariant. Hence, we conclude that the quasibosonic system with sub-bosonic harmonic Hamiltonian $\bar{H}$ will also be parity-invariant. Particularly, the eigenstates of this system will have definite parity just as the original sharp harmonic Hamiltonian. While all symmetries induced by LU operations on $H$ are inherited by $\bar{H}$, an interesting symmetry breaking may occur when time reversal (reversal of motion) is considered.

Whenever $\mathcal{O}$ is AA, one gets
\begin{equation}
\mathcal{O}\bar{a}_{\omega}\mathcal{O}^{-1}=\left(I_{0}^{*}+\frac{I_{1}^{*}}{2}\right)\mathcal{O}\hat{a}_{\omega}\mathcal{O}^{-1}+\frac{I_{1}^{*}}{2}\mathcal{O}\hat{a}^{\dagger}_{\omega}\mathcal{O}^{-1}.
\end{equation}
For $I_{0},I_{I} \in \mathbb{R}$ we obtain the the same conditions as in the LU case, if $\mathcal{O}\hat{a}_{\omega}\mathcal{O}^{-1}=\alpha\hat{a}_{\omega}$. 
Otherwise, we note that $\bar{a}_{\omega}$ will transform differently from $\hat{a}_{\omega}$.
The Hamiltonian $H=\hbar \omega(\hat{a}^{\dagger}_{\omega}\hat{a}_{\omega}+1/2)$ is invariant under time-reversal. However, when we go to the sub-bosonic Hamiltonian, 
$\bar{H}=\hbar \omega(\bar{a}^{\dagger}_{\omega}\bar{a}_{\omega}+{\cal C}/2)$,
we see that it is no longer symmetric under time-reversal if the coefficients $I_{0},I_{1}$ are complex. This means that, when changing from a bosonic commutation relation to a quasibosonic one, we have broken the time-reversal symmetry of the system, in general.
\subsection{Sub-bosonic Displacements and Coherent States}
At this point it is natural to consider deformed phase-space displacements \cite{barnett,recamier} as defined by
\begin{equation}
\label{displ}
\bar{D}(z)=e^{z\bar{a}_{\omega}^{\dagger}-z^{*}\bar{a}_{\omega}},
\end{equation}
where $z$ is an arbitrary complex number. That this is still a proper displacement operator is clear from its effect on the operator $\bar{a}_{\omega}$:
\begin{equation}
\bar{D}^{\dagger}(z)\bar{a}_{\omega}\bar{D}(z)=\bar{a}_{\omega}+[z^{*}\bar{a}_{\omega}-z\bar{a}_{\omega}^{\dagger},\bar{a}_{\omega}]=\bar{a}_{\omega}+{\cal C} z,
\end{equation}
where we used the Baker-Hausdorff fromula and the fact that $[\bar{a}_{\omega},\bar{a}_{\omega}^{\dagger}]$ is a scalar and, thus, commutes with $\bar{a}_{\omega}$ and $\bar{a}_{\omega}^{\dagger}$.
We can take a step further by showing that $\bar{D}(z)$, in fact, corresponds to a {\it standard} (sharp) displacement operator with a rescaled and rotated phase-space. To see that, we simply plug Eq. (\ref{fuzzya}) and its Hermitian conjugated into Eq. (\ref{displ}) and get
\begin{equation}
\bar{D}(z)=e^{\frak{z}\hat{a}_{\omega}^{\dagger}-\frak{z}^{*}\hat{a}_{\omega}}=\hat{D}(\frak{z}), \,\frak{z}=(I_0+I_1/2)z-(I_1/2)z^{*}.
\end{equation}
This relation, in turn, allows us to define sub-bosonic coherent states as $|\overline{z}\rangle=\bar{D}(z)|\overline{0}\rangle=\hat{D}(\frak{z})|\overline{0}\rangle$. By employing the coherent-state resolution of unity we get
\begin{equation}
|\overline{z}\rangle=\int\frac{d^2z'}{\pi} \langle z' |\hat{D}(\frak{z})|\bar{0}\rangle |z'\rangle=\int\frac{d^2z'}{\pi} \langle z'-\frak{z}|\bar{0}\rangle  |z'\rangle.
\end{equation}
Therefore,
\begin{equation}
|\overline{z}\rangle=\int\frac{d^2z'}{\pi}\, e^{-\frac{|z'-\frak{z}|^2}{2}}\, \left( \sum_k  \frac{\alpha_{2k}({z'}^*-\frak{z}^*)^{2k}}{\sqrt{(2k)!}}\right) |z'\rangle,
\end{equation}
where $\alpha_{2k}$ is given in Eq. (\ref{alpha}). We note that, as is the case of other non-canonical coherent states \cite{klauder,per}, the previous expression not necessarily coincides with
the eigenstates of $\bar{a}_{\omega}$, that is, the coherent sum
\begin{equation}
||\overline{z}\rangle\rangle=e^{-\frac{|z|^2}{2}}\sum_n\frac{z^n}{\sqrt{n!}}|\overline{n}\rangle\ne |\overline{z}\rangle ,
\end{equation}
which constitutes an alternative way to define sub-bosonic coherent states.
\section{Quantization of quasi-bosonic fields}
\label{qft}
We proceed by considering a free bosonic field, whose energy density can be written in terms of the ``quadrature'' fields as
\begin{equation}
\nonumber
{\mathcal H}=\frac{1}{2}\sum_{\mathbf{k},\chi}(\hat{P}^2_{\mathbf{k},\chi}+\omega^2\hat{Q}^2_{\mathbf{k},\chi}),
\end{equation}
where $\chi$ refers to internal degrees of freedom such as spin and polarization. By ``free'' we mean that no other fields are present or, at least, do not interact with the bosons under study. However, often in quantum field theory, assuming that other elementary fields are absent is equivalent to say that their vacua (ground states) are {\it present}. Thus, rigorously speaking, one cannot always get rid of the other fields, even in principle. In these cases, the hypothesis of a free field is an approximation.

We consider the possibility that vacua related to these other fields have some tangible effect on the field under consideration, as is the case of the electron field in the electromagnetic vacuum. In trying to take this into account, we will assume that the second-quantization creation and annihilation operators inherit the commutation relation (\ref{comm2}). Loosely speaking, we will assume that some factor blurs the bosonic field under consideration. Of course, this is not a rigorous argument and we may simply state that our goal is to investigate the consequences of the previous assumption, at least on formal grounds.
We therefore write
\begin{equation}
\nonumber
\bar{\mathcal H}=\sum_{\mathbf{k}}\bar{H}_{\mathbf{k}}=\sum_{\mathbf{k}}\left[\hbar \omega\bar{a}_{\mathbf{k}}^{\dagger}\bar{a}_{\mathbf{k}}+\frac{\hbar \omega}{2} \mathcal{C}(\omega)\right],
\end{equation}
with an arbitrary eigenstate given by
\begin{equation}
\label{prod}
\nonumber
 \prod_{\mathbf{q=q}_1}^{\mathbf{q}_f}\frac{(\bar{a}_{\mathbf q}^{\dagger})^{n_{\mathbf q}}}{\sqrt{{n_{\mathbf q}}!}}|\overline{0}\rangle=|\cdots,\bar{n}_{\mathbf q_{j-1}},\bar{n}_{\mathbf q_j}, \bar{n}_{\mathbf q_{j+1}}, \cdots\rangle,
\end{equation}
where $\mathbf{q}$ is a collective index, including momentum and intrinsic degrees of freedom.
Due to (\ref{number}) the energy of a single excitation of a mode with angular frequency $\omega$ is 
\begin{equation}
\nonumber
\langle \bar{n}|\bar{H}|\bar{n}\rangle-\langle \overline{n-1}|\bar{H}|\overline{n-1}\rangle=\mathcal{C}(\omega)\hbar\omega\equiv \Delta \bar{E}(\omega),
\end{equation}
leading, generally, to nonlinear dispersion-like relations, due to the $\omega$ dependence of the commutation function.
In the particular case of the natural-line width distribution we have
\begin{equation}
\nonumber
\Delta\bar{E}(\omega)=\frac{\hbar\omega}{\sqrt{1+\zeta(\omega)^2}},
\end{equation}
\begin{figure}
\label{f2}
\centering
\resizebox{0.75\textwidth}{!}{
 \includegraphics{{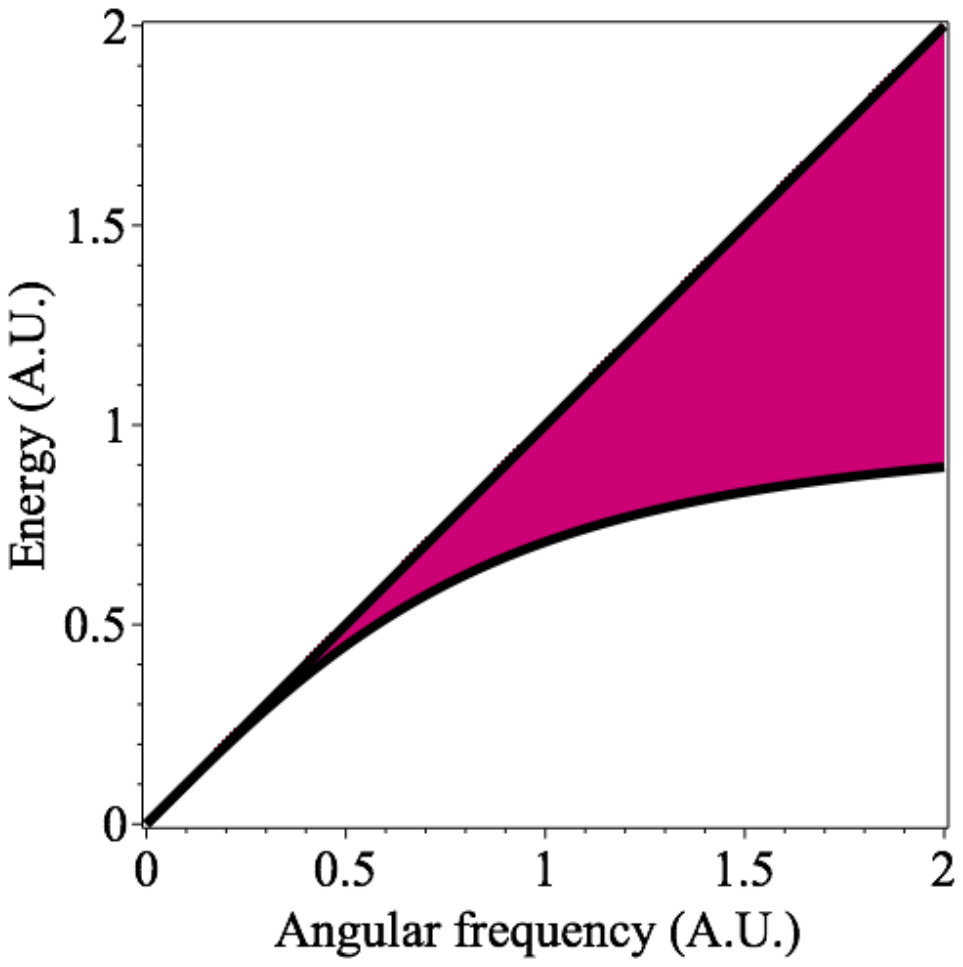}}
}
\caption{The upper curve represents $\Delta\bar{E}(\omega)=\rm{const} \times \hbar \omega$ and the lower curve $\Delta\bar{E}(\omega)=\hbar\omega/\sqrt{1+\rm{const}'\times\omega^2}$,
in arbitrary units (A.U.) with $\hbar=1$, const$=$const$'=1$. Any $\Delta\bar{E}(\omega)$ satisfying the conditions specified in the text which, in addition, have a monotonous first derivative, 
must remain in the filled area.}
\end{figure}
whose dependence on the frequency mode is determined by $\zeta(\omega)=\Gamma(\omega)/2\omega$, that is, on how the typical width associated with the sub-bosonic operators depends
on $\omega$. If one demands that $\Delta\bar{E}$ remains finite as $\omega \rightarrow 0$, we must assume that $\Gamma(\omega) \sim \omega^{\mu}$ as $\omega \rightarrow 0$ 
with $\mu >1$. On the other hand, if we are to retain the property that $\Delta\bar{E}(\omega)$ is a monotonically increasing function of $\omega$, we must have $1+\zeta^2<c\omega^2$,
or $\Gamma(\omega)< \omega\sqrt{c\omega^2-1}$, where $c$ is a positive, otherwise arbitrary constant. In the limit of large $\omega$ we get $\Gamma(\omega)\sim \omega^{\mu'}$, with
$\mu'\le 2$. As an illustration, let us consider the two limiting behaviors $\Gamma(\omega)\propto \omega$ and $\Gamma(\omega)\propto \omega^2$. 
This would lead to $\Delta\bar{E}(\omega)={\rm const} \times \hbar \omega$ and $\Delta\bar{E}(\omega)=\hbar\omega/\sqrt{1+{\rm const}'\times\omega^2}$, respectively. These two limiting functional forms are depicted in figure 2. Any intermediate relation satisfying the previous conditions with a monotonic first derivative must lay in the filled area.

\section{Final Remarks}
Typically, deformed algebras involving ladder operators lead to (or start from) commutation relations between annihilation and creation which depend on phase-space variables.
In the present work, the dependence of creation and annihilation operators on the frequency is formally exploited to define sub-bosonic operators. Based on this notion, we find a commutation relation which reads $[\bar{a}_{\omega},\bar{a}_{\omega}^{\dagger}]= {\cal C}$, where ${\cal C}$ is a scalar ($\ne 1$ in general). 
Although we characterized the basic features o sub-bosonic systems, in-deep studies in the context of elementary quantum mechanics and quantum statistical mechanics are desirable. 
In what concerns QFT, the dependence of ${\cal C}$ on the mode may bring a relevant ingredient, namely, the commutation relation induces a nonlinear dispersion-like relation, even in the case
of free fields.

The authors thank Azadeh Mohammadi, Bruno Cunha, and Carlos Batista for their comments on an early version of this work. This research received financial support from the Brazilian agencies Coordena\c{c}\~ao de Aperfei\c{c}oamento de Pessoal de N\'{\i}vel Superior (CAPES), Funda\c{c}\~ao de Amparo \`a Ci\^encia e Tecnologia do Estado de Pernambuco (FACEPE), and Conselho Nacional de Desenvolvimento Cient\'{\i}fico  e Tecnol\'ogico through its program CNPq INCT-IQ (Grant 465469/2014-0). R.X. is also supported by the U.S. Department of Energy under the contract DE-SC0017647.

\section{Appendix: Complex Integration}
In this appendix we evaluate the functional
\begin{equation}
I_{k}[f']=\int^{\infty}_{-\infty}\frac{x^k f'(x)}{\sqrt{x+1}} d{x}.
\end{equation}
With the substitution $x+1 \rightarrow x$ we get
\begin{equation}
I_{k}[f']=\int^{\infty}_{-\infty}\frac{(x-1)^k f'(x-1)}{\sqrt{x}}=\int^{\infty}_{-\infty}\frac{g_{k}(x)}{\sqrt{x}} d{x}.
\end{equation}
where $g_{k}(x)=(x-1)^k f'(x-1)$. We start by considering the following integration in the complex plane
\begin{equation}\label{compint}
I^{\gamma}_{k}[g]=\int_{\gamma}\frac{g_{k}(z)}{\sqrt{z}} d{z}.
\end{equation}
Note that $\sqrt{z}$ has branch points at zero and at infinity. For the remainder of this appendix, we will work with the branch cut connecting these two branch points through the negative imaginary axis, that is, we restrict ourselves to $\theta \in [e^{\pi i/2},e^{3 \pi i/2})$. We will assume the following condition on the function $g_{k}(z)$:

\begin{itemize}
\item Condition (i): $g_{k}(z)$ is analytic in ${\rm H} \cup \mathbb{R}$, except for a finite number of removable singularities (poles) in ${\rm H}$. Here, ${\rm H}\equiv\left\lbrace z\in \mathbb{C} ~|~ \Im{(z)}> 0 \right\rbrace$ is the upper-half of the complex plane.
\end{itemize}

Given that this condition is satisfied, let us evaluate expression (\ref{compint}) on the path $\gamma$ defined as
\begin{equation}
\gamma=\gamma_{-}+\gamma_{\epsilon}+\gamma_{+}+\gamma_{R},
\end{equation}
with the following parametrizations:
\begin{align}
\nonumber
\gamma_{-}(r)&=re^{i\pi}, ~ r \in (R,\epsilon], \quad \gamma_{\epsilon}(\theta)=\epsilon e^{i\theta}, ~ \theta \in [0,\pi], \\ 
\nonumber
\gamma_{+}(r)&=r,~ r \in [\epsilon,R),~~~~ \quad \gamma_{R}=Re^{i \theta}, ~ \theta\in[\pi, 0],
\end{align}
where we will take the limits $R \to \infty$ and $\epsilon \to 0$. Let us now evaluate each of the integrals
\begin{eqnarray}
\nonumber
I^{{\gamma_{-}}}_k[g]=\int_{\gamma_{-}}\frac{g_{k}(z)}{\sqrt{z}} d{z}= \lim_{\substack{\epsilon \to 0 \\ R \to \infty}}\int^{\epsilon}_{R}\frac{g_{k}(-r)}{\sqrt{re^{i\pi}}} d{(-r)}\\
=\int^{0}_{\infty}\frac{g_{k}(-r)}{\sqrt{-r}} d{(-r)}\underset{(-r) \to x}{=} \int^{0}_{-\infty}\frac{g_{k}(x)}{\sqrt{x}} d{x}
\end{eqnarray}
\begin{eqnarray}
\nonumber
I^{\gamma_{+}}_k[g]=\int_{\gamma_{+}}\frac{g_{k}(z)}{\sqrt{z}} d{z}= \lim_{\substack{\epsilon \to 0 \\ R \to \infty}}\int^{R}_{\epsilon}\frac{g_{k}(r)}{\sqrt{r}} d{r}\\
\underset{r \to x}{=} \int^{\infty}_{0}\frac{g_{k}(x)}{\sqrt{x}} d{x}
\end{eqnarray}
\begin{eqnarray}
\nonumber
I^{\gamma_{R}}_k[g]=\int_{\gamma_{R}}\frac{g_{k}(z)}{\sqrt{z}} d{z}= \lim_{\substack{R \to \infty}}\int_{\gamma_{R}}\frac{g_{k}(Re^{i\theta})}{\sqrt{Re^{i\theta}}} d{(R e^{i\theta})}\\
=\lim_{\substack{R \to \infty}}i\int^{\pi}_{0}\frac{g_{k}(Re^{i\theta})}{\sqrt{e^{i\theta}}} R^{1/2}d{\theta}
\end{eqnarray}
\begin{eqnarray}
\nonumber
I^{\gamma_{\epsilon}}_k[g]=\int_{\gamma_{\epsilon}}\frac{g_{k}(z)}{\sqrt{z}} d{z}= \lim_{\substack{\epsilon \to 0}}\int^{0}_{\pi}\frac{g_{k}(\epsilon e^{i\theta})}{\sqrt{\epsilon e^{i\theta}}} d{(\epsilon e^{i\theta})}\\
=\lim_{\substack{\epsilon \to 0}}i\int^{\pi}_{0}\frac{g_{k}(Re^{i\theta})}{\sqrt{e^{i\theta}}} \epsilon^{1/2}d{\theta}
\end{eqnarray}
Now, we also need to assume the following condition on $g(z)$:
\begin{itemize}
\item Condition (ii): We associate with $g_{k}(z)$ the coefficients $\alpha,\beta$ defined as
\begin{equation}
g_{k}(Re^{i\theta}) \underset{R \to \infty}{\sim} R^\alpha, \quad \text{and} \quad g_{k}(\epsilon e^{i\theta}) \underset{\epsilon \to 0}{\sim} \epsilon^\beta.
\end{equation}
These coefficients must be such that $\alpha <-1/2$ and $\beta >-1/2$.
\end{itemize}
If condition (ii) is satisfied, we can set
\begin{equation}
I^{\gamma_{R}}_k [g]\underset{R \to \infty}{\to} 0, \quad \text{and} \quad I^{\gamma_{\epsilon}}_{k}[g]\underset{\epsilon \to 0}{\to} 0. 
\end{equation}
Combining these results, Eq.\eqref{compint} gives
\begin{equation}\label{compint1}
I^{\gamma}_{k}[g]=I^{\gamma_{-}}_k[g]+I^{\gamma_{\epsilon}}_{k}[g]+I^{\gamma_{+}}_{k}[g]+I^{\gamma_{R}}_{k}[g]=\int^{\infty}_{-\infty}\frac{g_{k}(x)}{\sqrt{x}} d{(x)}.
\end{equation}
However, we can also apply the residue theorem to Eq.\eqref{compint}:
\begin{equation}\label{compint2}
I^{\gamma}_{k}[g]=2 \pi i \sum_{a \in H}\underset{z=a}{\operatorname{Res}}g(z).
\end{equation}
Hence, equating Eq.\eqref{compint1} and Eq.\eqref{compint2}, we get
\begin{equation}
\int^{\infty}_{-\infty}\frac{g_{k}(x)}{\sqrt{x}} d{x}=2 \pi i \sum_{a \in H}\underset{z=a}{\operatorname{Res}}~\frac{g_{k}(z)}{\sqrt{z}},
\end{equation}
which ends the proof of proposition 1.

\end{document}